\documentclass[12pt,amssymb]{article}

\let\d=\delta
\let\e=\epsilon

\newcommand{\be}{\begin{equation}}
\newcommand{\ee}{\end{equation}}
\newcommand{\bea}{\begin{eqnarray}}
\newcommand{\eea}{\end{eqnarray}}

\newcommand{\del}{\partial}

\begin{document}

\begin{titlepage}
\begin{center}
\vskip .2in \hfill \vbox{
    \halign{#\hfil         \cr
           hep-th/0310056 \cr
           MCTP-03-43  \cr
           UCSD/PTH 03-13 \cr
           UCI-2003-44 \cr
           October 2003    \cr
           }  
      }   
\vskip 1.5cm {\Large \bf Supergravity Solutions For $AdS_3 \times S^3$ Branes} \\
\vskip .1in \vskip .3in {\bf  Jason Kumar}
\footnote{e-mail address:jpkumar@umich.edu}\\
\vskip .15in {\em  Michigan Center for Theoretical Physics,
University of Michigan\\
Ann Arbor, MI  48105 USA \\} \vskip .2in {\bf Arvind Rajaraman}
\footnote{e-mail address:arajaram@uci.edu} \vskip .15in {\em
Department of Physics,
University of California, Irvine \\
Irvine, CA 92697\\} \vskip .1in \vskip 1cm
\end{center}
\begin{abstract}
We find a large class of supersymmetric solutions in the $AdS_3
\times S^3$ background with NS fluxes. This two-parameter family
of solutions preserves 8 of the 16 supersymmetries of the
background. \vskip 0.5cm
\end{abstract}
\end{titlepage}
\newpage

\section{Introduction}

In recent years, a great deal of progress has been made in the
study of both gauge theory and gravity through the AdS/CFT
correspondence.  This correspondence allows one to relate a theory
of quantum gravity in a particular background to a dual conformal
field theory living on the boundary.  This correspondence is best
understood when the bulk space-time background is an anti-de
Sitter space.  Unfortunately, even in this case, it is quite
difficult to study the bulk gravitational theory beyond the
supergravity limit.  Recently, some progress has been made in this
area in several different contexts (\cite{Berenstein:2002jq}
\cite{Bena:2003wd}, among others).

However, the best understood case is the $AdS_3 \times S^3 \times
T^4 $ solution with NS-NS gauge fields turned on.  This background
arises from the embedding of a stack of fundamental strings within
a stack of NS-5-branes (when one then takes the near-horizon
limit).  The action of a string in this background is a
Wess-Zumino-Witten model on the group manifold $SL(2,R)\times
SU(2)$. Since $SL(2,R)$ is non-compact, this theory is still
somewhat subtle; nevertheless progress has been made in studying
both the open and closed strings in this background
(\cite{mo}\cite{Rajaraman:2001cr}).

It is thus useful to construct supergravity solutions for D-branes
in $AdS_3$, which will then be described in a dual description by
a field theory. In the particular case that we shall analyze, the
supergravity solution we find is believed to be dual to a defect
conformal field theory on the boundary \cite{Karch:2000ct}.

We will be interested in defects which preserve 8 of the 16
supersymmetries of the $AdS_3 \times S^3 \times T^4 $ background.
In \cite{Bachas:2000fr}, it was shown using boundary state
arguments that a D3-brane embedded in an $AdS_2 \times S^2$
submanifold of this background is ${1\over 2}$ BPS. This is twice
the number preserved by a system containing D3-branes, fundamental
strings and NS-5 branes, due to the enhancement of supersymmetry
\cite{Kallosh:1997qw}.  General conditions for the preservation of
supersymmetry by branes in $AdS_3$ were also considered by
\cite{Giveon:2003ku}.

Supergravity solutions for general intersecting branes were
considered in \cite{Rajaraman:2002vf,Rajaraman:2000dn,
Rajaraman:2000ws,Fayyazuddin:2000em,Brinne:2000nf,Brinne:2000fh,Cherkis:2002ir,
Lunin:2001jy}.
Secondly, in \cite{Bachas:2000ik,Sarkissian:2003jn}, brane probes
were considered.
The equations of motion derived from the Born-Infeld action were
then solved to produce stable supersymmetric branes.
Unfortunately, the latter approach only produces solutions to
linearized supergravity. Instead, we will follow the general
approach of \cite{Rajaraman:2000dn,Fayyazuddin:2000em}.  This
generalizes the result of \cite{Kumar:2002wc}.

In the section 2, we study the $\kappa$-symmetry of D3-branes in
$AdS_3 \times S^3$.  The gauge-fixing of $\kappa$-symmetry will
determine which Killing spinors are preserved by the supergravity
solution.  In section 3, we use this to solve the Killing
equations.  In \cite{Kumar:2002wc}, a single ${1\over 2}$-BPS
solution was found.  This restriction on the location of the
sources was due to certain assumptions made in order to make the
Killing equations more tractable.  Using much more general
assumptions, we are able to find a solution written in terms of
one function, with two non-linear partial differential equations
as constraints.  We argue that a two-parameter family of solutions
exists.  Generically, the sources for these solutions do not
appear to be localized (at least to first order).

\section{Obtaining Supersymmetry from $\kappa$-Symmetry}

A D-brane
typically breaks some of the supersymmetries of the background.
The preserved supersymmetries (labelled by Killing spinors) are
most easily found by the analysis of $\kappa$-symmetry.

In a consistent supersymmetric brane solution, the preserved Killing
spinors (when projected onto the brane) must be invariant under a gauged
fermionic symmetry known as $\kappa$-symmetry. The projector which
projects onto these invariant spinors can be found
very generally, and from this we can deduce the form of the preserved
Killing spinors.

The background metric of $AdS_3 \times S^3$ in global coordinates
may be written as \bea ds^2 =d\psi^2 +\cosh^2 \psi(d\omega^2
-\cosh^2 \omega d\tau^2 )+
\nonumber\\
d\theta^2 +\sin^2 \theta (d\phi^2 + \sin^2 \phi d\chi^2 )
\eea
The background $B_{NS}$ fields are
\bea
\bar{B}_{\tilde \phi \tilde \chi} &=&
{1\over 2} (\theta - {\sin 2\theta \over 2}) \sin \phi
\nonumber\\
\bar{B}_{\tilde \tau \tilde \omega} &=& {1\over 2} (\psi - {\sinh 2\psi \over
2}) \cosh \omega
\eea

This background preserves a 16-component Killing spinor.

In \cite{Bachas:2000fr}, it was argued that a D3-brane embedded
along an $AdS_2 \times S^2$ submanifold of $AdS_3 \times S^3$
would preserve 8 of the 16 supersymmetries of the background. This
embedding may be parameterized as a brane stretching along the
($\tau$, $\omega$, $\phi$, $\chi$) coordinates of the global
coordinate system, with the $\psi$ and $\theta$ coordinates acting
as transverse scalars.

One needs to turn on a gauge field strength on the D3-brane, which
is of the form \bea 4\pi \alpha^{'} F_{\phi \chi} &=& -\pi p\sin
\phi
\nonumber\\
4\pi \alpha^{'} F_{\tau \omega} &=& -\pi q\cosh \omega \eea

The Born-Infeld Lagrangian for this D3-brane is then \bea L_{DBI}
&=& -T \sqrt{-\det M} \eea where \bea \sqrt{-\det M} &=&N(\psi)
L(\theta) \cosh \omega \sin \phi
\nonumber\\
L(\theta) &=& (\sin^4 \theta +(\theta -{\sin 2\theta \over 2} -\pi
p)^2 )^{1\over 2}
\nonumber\\
N(\psi) &=& (\cosh^2 \psi -(\psi +{\sinh 2\psi \over 2} -\pi q )^2
)^{1\over 2} \eea

The resulting equations of motion are solved by

\bea \theta_0 &=& \pi p
\nonumber\\
\psi_0 &=& \pi q \eea and we have \bea L(\theta_0 ) &=& \sin
\theta_0
\nonumber\\
N(\psi_0 ) &=& \cosh \psi_0
\nonumber\\
{\cal F}_{\phi \chi} &=& -{1\over 2} \sin 2\theta_0 \sin \phi
\nonumber\\
{\cal F}_{\tau \omega}&=& -{1\over 2} \sinh 2\psi_0 \cosh \omega
\eea

The projection onto $\kappa$-invariant spinors is given
by \cite{Cederwall:1996pv}

\be d^{p+1} \xi \Gamma_0 = -e^{-\Phi} L_{DBI} ^{-1} e^{F^{total}}
\wedge X|_{vol}, \ee

where

\bea X &=& \bigoplus_n \Gamma_{(2n)} K^n I
\nonumber\\
K\psi &=& \psi^*
\nonumber\\
I\psi &=& -\imath \psi
\nonumber\\
\Gamma_{(n)} &=& {1\over n!} d\xi^{i_n} \wedge ... \wedge
d\xi^{i_1} \Gamma_{i_1 ... i_n} \eea

We see that

\bea \Gamma_0 &=& -\imath (\cos \theta_0 \cosh \psi_0 \gamma_{\tau
\omega} K + \sin \theta_0 \cosh \psi_0 \gamma_{\tau \omega \phi
\chi}
\nonumber\\
&+&\sin \theta_0 \sinh \psi_0 \gamma_{\phi \chi} K +\cos \theta_0
\sinh \psi_0  ) \eea where $\Gamma_0$ is traceless and $\Gamma_0
^2 =1$.

The $\kappa$-symmetry projector $\Gamma_0$ acts on
the background killing spinor at the brane. We also need the
projector away from the brane.
Let $\e$ be the Killing spinor preserved in the presence of the
brane. We will assume that $\epsilon $ is of the form $ \epsilon =
f \bar{\epsilon} $ where $f$ is a complex function, and
$\bar{\epsilon} $ is the Killing spinor of the background $AdS_3
\times S^3$.

\bea \bar \epsilon &=& \exp\left({- \psi\over 2} \gamma_{\tau
\omega} K\right) \exp\left({- \theta \over 2} \gamma_{\phi \chi}
K\right) R_0 (\phi ,\chi ,\omega ,\tau) \bar \epsilon_0
\nonumber\\
&=& \Lambda \bar \epsilon (\psi_0 ,\theta_0 ) \eea where \bea
\Lambda &=& \exp\left({-(\psi-\psi_0 )\over 2} \gamma_{\tau
\omega} K\right) \exp\left({-(\theta -\theta_0 ) \over 2}
\gamma_{\phi \chi} K\right) \eea

To preserve supersymmetry, we must have
$(1-\Gamma_0)\bar \e=0$ at $\psi=\psi_0$
and $\theta=\theta_0$ (but at arbitrary $\tau$, $\omega$, $\phi$
and $\chi$ ), i.e. $(1-\Gamma_0)\bar{\e}(\psi_0,\theta_0)=0$.

We can write this globally as $(1-\Gamma)\e=0$ where

\bea \Gamma = f\Lambda \Gamma_0 \Lambda^{-1}  f^{-1} \eea After
some algebra, we find \bea \Gamma = -\imath (M +N \gamma_{\tau
\omega} K +O \gamma_{\phi \chi }K +P\gamma_{\tau \omega \phi
\chi})
\nonumber\\
M=\cos \theta \sinh \psi \qquad  N = {f\over f^*}\cos \theta \cosh
\psi
\nonumber\\
O ={f\over f^*} \sin \theta \sinh \psi \qquad  P = \sin \theta
\cosh \psi \eea

Note that this projector is now independent of $\psi_0$ and
$\theta_0$. We can rewrite this projector in the form

\bea \label{eq15} \gamma_{\tau \omega} K \epsilon = (A +B
\gamma_{\psi \theta}) \epsilon \eea where \bea  A &=& -{f^* \over
f} {\cosh \psi \over \cosh^2 \psi -\sin^2 \theta} (\sinh
\psi+\imath \cos \theta)
\nonumber\\
B &=& {f^* \over f} {\imath \sin \theta \over \cosh^2 \psi -\sin^2
\theta} (\sinh \psi +\imath \cos \theta)
 \eea

\section{The Killing Equations}

We have found the form of the supersymmetry. We now turn to a more
detailed analysis of the supersymmetry
\footnote{We use the conventions of \cite{Kumar:2002wc}, which arise
from \cite{Bergshoeff:1996wk} and \cite{Kallosh:1997jz} }.

The supersymmetry variations are generally of the form \bea
\d\psi_\mu=D_\mu\e+... \eea Supersymmetry is preserved if the
variations all vanish. Every choice of $\e$ for which these
variations vanish is called a Killing spinor, and produces a
different preserved supersymmetry.

In our case we have already found the form of the Killing spinor.
We are thus requiring that the supersymmetry variations vanish for
the Killing spinors of the form $ \epsilon = f \bar{\epsilon} $,
with $(1-\Gamma)\e=0$.

We will now explicitly write out the form of the supersymmetry
variations, substitute the form of the Killing spinor into them,
and find the background solution that preserves this Killing
spinor.

 We begin with a few
assumptions to constrain the solution. First, we assume that this
solution is self-dual in 6 dimension, ie. $G^{\psi \tau \omega} =
G^{\phi \chi \theta}$ and
$G^{\tau \omega \theta} =-G^{\psi \phi \chi}$.  The killing equations
(for the dilatino) then imply that
complex axion-dilaton $\Phi$ is constant.
Furthermore, the equations of motion and generalized Bianchi
identity imply that the 5-form field strength $F$ vanishes.
We also assume that all fields depend only on the coordinates
$\psi$ and $\theta$.

The killing equations in Einstein frame can be written as
\cite{Duff:1992hu}
\bea
\del_{\tilde \mu}\e +{1\over 4} \omega_{\tilde \mu} ^{ab}
\gamma_{ab} +{\imath \over 192}  F_{\tilde \mu}^{~bcde}
\gamma_{bcde}\epsilon^* +{\imath \over 48} e^{\Phi} (G^{abc}
\gamma_{\tilde \mu abc}- 9G_{\tilde \mu}^{~ab} \gamma_{ab})
\epsilon^*=0
\eea
where $G$ is a complex field whose real and imaginary parts
are the RR and NS-NS three-form field strengths, respectively.
The value of this field for the background is
$\bar{G}^{\psi \tau \omega}=\bar{G}^{\phi \chi \theta} =-\imath$.

For example, the $\psi$ Killing equation can be written as \bea
-{\partial_{\psi} f \over f }\gamma_{\psi} \epsilon -{1\over 2}
\omega_{ \psi} ^{\psi \theta} \gamma_{ \theta} \epsilon +{\imath
\over 2}  G^{\psi\tau \omega} \gamma_{\psi\tau \omega} \epsilon^*
- {\imath \over 2}  G ^{\tau\omega \theta} \gamma_{\tau\omega
\theta} \epsilon^*  =
\nonumber\\
{1 \over 2}f {e_{\psi } ^{\tilde \psi} \over \bar{e}_{\psi }
^{\tilde \psi}} \gamma_{\psi\tau \omega} \bar\epsilon^* \eea and
the other Killing equations take a similar form.

Since we know that $\e$ satisfies the $\kappa$-symmetry
projection, we can use (\ref{eq15}) to simplify these equations.
The above equation, for example becomes : \bea -{\partial_{\psi} f
\over f }\gamma_{\psi} \epsilon -{1\over 2} \omega_{ \psi} ^{\psi
\theta} \gamma_{ \theta} \epsilon +{\imath \over 2}  G^{\psi\tau
\omega} (A\gamma_{\psi} +B\gamma_{\theta}) \epsilon + {\imath
\over 2}  G ^{\tau\omega \theta} (B\gamma_{\psi} -A\gamma_{\theta
}) \epsilon  =
\nonumber\\
{1 \over 2}{f \over f^*} {e_{\psi } ^{\tilde \psi} \over
\bar{e}_{\psi } ^{\tilde \psi}} (A\gamma_{\psi } +B\gamma_{\theta
}) \epsilon \eea

This is only satisfied if the coefficients of $\gamma_{\psi}$ and
$\gamma_{ \theta}$ are each zero. Therefore we get two complex
equations. The other Killing equations similarly yield other
complex equations.

By taking linear combinations of these equations, we solve for the
various field strengths in terms of the spin connections, and also
get equations for the spin connections.
The solution for the field strengths in terms of the spin
connection is then \bea  -\imath e^{\Phi}G^{\tau \omega \theta}
(A^2 + B^2) = {1\over 2} \omega_{\omega} ^{\omega \psi }B -{1\over
2} \omega_{\omega} ^{\omega \theta}A -{1\over 2} \omega_{\phi}
^{\phi \psi}B +{1\over 2} \omega_{\phi} ^{\phi \theta}A
\nonumber\\
+{1\over 2}A \cot \theta {e_{\phi} ^{\tilde \phi } \over
\bar{e}_{\phi } ^{\tilde \phi }} +{1\over 2}B \tanh \psi
{e_{\omega} ^{\tilde \omega } \over \bar{e}_{\omega } ^{\tilde
\omega }} \nonumber\\
G^{\psi \tau \omega} = \imath \beta G^{\tau \omega \theta} -\imath
{e_{\psi } ^{\tilde \psi } \over \bar{e}_{\psi } ^{\tilde \psi }}
({f\over f^*} +{\del_{\tilde \psi } \log f^2 \over A}) \eea

We also have the constraints\bea Re \log f^2 = \log {e_{\phi}
^{\tilde \phi } \over \bar{e}_{\phi } ^{\tilde \phi }} =\log
{e_{\omega} ^{\tilde \omega
} \over \bar{e}_{\omega } ^{\tilde \omega }} ~~~~~~~~~~~~~
~~~~~~~~~~~~~\nonumber\\
\nonumber\\-Im {\del_{\psi} f\over f} = {1\over \beta}[-{1\over 2}
\omega_{\psi} ^{\psi \theta} +{1\over 2} \omega_{\phi} ^{\phi
\theta} +{1\over 2} \cot \theta {e_{\phi} ^{\tilde \phi } \over
\bar{e}_{\phi } ^{\tilde \phi }}]~~~~~~~~~~~~~
\nonumber\\
-Im {\del_{\theta } f\over f} =\beta[{1\over 2} \omega_{\theta}
^{\theta \psi } -{1\over 2} \omega_{\omega} ^{\omega \psi }
-{1\over 2} \tanh \psi {e_{\omega} ^{\tilde \omega } \over
\bar{e}_{\omega } ^{\tilde \omega }}]~~~~~~~~~~~~~ \nonumber\\
\label{eq22} -{e_{\theta} ^{\tilde \theta } \over \bar{e}_{\theta
} ^{\tilde \theta }}
\partial_{\tilde \theta} \log
({e_{\tilde \omega} ^{\omega } \over \bar{e}_{\tilde \omega}
^{\omega }} {e_{\tilde \phi } ^{\phi } \over \bar{e}_{\tilde \phi
} ^{\phi }}) +({e_{\phi } ^{\tilde \phi } \over \bar{e}_{\phi }
^{\tilde \phi }} -{e_{\theta} ^{\tilde \theta } \over
\bar{e}_{\theta } ^{\tilde \theta }}) \cot \theta = 0~~~~~~~~~~~~~
\nonumber\\
-{e_{\psi} ^{\tilde \psi } \over \bar{e}_{\psi } ^{\tilde \psi }}
\partial_{\tilde \psi } \log
({e_{\tilde \omega} ^{\omega } \over \bar{e}_{\tilde \omega}
^{\omega }} {e_{\tilde \phi } ^{\phi } \over \bar{e}_{\tilde \phi
} ^{\phi }}) +({e_{\omega } ^{\tilde \omega } \over
\bar{e}_{\omega } ^{\tilde \omega }} -{e_{\psi } ^{\tilde \psi }
\over \bar{e}_{\psi } ^{\tilde \psi }}) \tanh \psi = 0~~~~~~~~~~~~~ \nonumber\\
  -{e_{\psi } ^{\tilde \psi } \over \bar{e}_{\psi }
^{\tilde \psi }}
\partial_{\tilde \psi } \log
({e_{\tilde \theta} ^{\theta } \over \bar{e}_{\tilde \theta}
^{\theta }} {e_{\omega } ^{\tilde \omega } \over \bar{e}_{\omega }
^{\tilde \omega }}) -{\beta^2 \over 1-\beta^2} ({e_{\psi }
^{\tilde \psi } \over \bar{e}_{\psi } ^{\tilde \psi }} -{e_{\omega
} ^{\tilde \omega } \over \bar{e}_{\omega } ^{\tilde \omega }})
\tanh \psi = {1\over 1-\beta^2} ({e_{\phi } ^{\tilde \phi } \over
\bar{e}_{\phi } ^{\tilde \phi }} -{e_{\theta } ^{\tilde \theta }
\over \bar{e}_{\theta } ^{\tilde \theta }}) \tanh \psi \eea

We  now define
\bea X = {e_{\tilde \theta } ^{\theta } \over
\bar{e}_{\tilde \theta } ^{\theta }} {e_{\omega } ^{\tilde \omega
} \over \bar{e}_{\omega } ^{\tilde \omega }} \qquad Y = {e_{\tilde
\psi } ^{\psi } \over \bar{e}_{\tilde \psi } ^{\psi }} {e_{\omega
} ^{\tilde \omega } \over \bar{e}_{\omega } ^{\tilde \omega }}
\qquad Z = {e_{\omega } ^{\tilde \omega } \over \bar{e}_{\omega }
^{\tilde \omega }} ={e_{\phi } ^{\tilde \phi } \over \bar{e}_{\phi
} ^{\tilde \phi }} \qquad W = {X\over Y} \eea

Using the constraints, we can eliminate $Z$, $X$, $Y$ in favor
of $W$, to get the equations:
\bea
{\cos^2 \theta -\sinh^2 \psi \over \cosh^2
\psi} (W -{1\over W}) = -{\cot \theta \over W^2} \del_{\tilde
\theta} W -\tanh \psi \del_{\tilde \psi} W \eea and \bea
\del_{\tilde \psi} \del_{\tilde \theta} W -{1\over W}\del_{\tilde
\psi} W \del_{\tilde \theta} W +{1\over 1-\beta^2}{1\over W}\tanh
\psi \del_{\tilde \theta} W
\nonumber\\
-{2\beta^2 \over (1-\beta^2)^2} \tanh \psi \cot \theta (1-W^2)
-{\beta^2 \over 1-\beta^2}W\cot \theta
\partial_{\tilde \psi} W =0
\eea

Any solution is determined by the function $W$, which must satisfy
the above two non-linear partial differential equations. The
constraint equations  then allow one to solve for $X$,$Y$ and $Z$.
With $W$, this solves for all of the vierbeins. The other
 equations then allow one to solve for $f$ and $G$, which
gives the full solution.

Note that the constant solution $W=1$ was the solution found in
\cite{Kumar:2002wc}.  Since the equations are non-linear, it is
difficult to find the most general consistent solution to both
equations.  The equations may be expanded around $W=1$, and we have
checked that, at least to third order in $(W-1)$, there exists a
one-parameter family of solutions. This suggests that the full
non-linear equations for $W$ in fact exhibits a consistent
one-parameter family of solutions, with $W=1$ being a point on
this line.

To linear order, we may write the solution for $W$ as \bea W =
1+a(\sinh^2 \psi +\cos^2 \theta ) \eea where $a$ is a constant of
integration. Solving the equations of the vierbein yields another
constant of integration, $b$.
We then find to linear order in $(W-1)$
\bea X = 1-{b\over \sin
\theta \cosh \psi} -a\sin^2 \theta
\nonumber\\
Y = 1-{b\over \sin \theta \cosh \psi} -a\cosh^2 \psi
\nonumber\\
Z = 1-{1\over 2}{b\over \sin \theta \cosh \psi} +{1\over
4}a(\sinh^2 \psi +\sin^2 \theta)
\nonumber\\
Re f^2 = Z
\nonumber\\
Im \log f = {1\over 2} a\sinh \psi \cos \theta
\nonumber\\
G^{\tau \omega \theta} =-G^{\psi \phi \chi} = {1\over 2}[{b\over
\sin^2 \theta \cosh^2 \psi} +a\sin \theta \cosh \psi +\imath
{b\cot \theta \tanh \psi \over \sin \theta \cosh \psi} ]
\nonumber\\
G^{\psi \tau \omega} = G^{\phi \chi \theta} = -\imath +{1\over
2}a\sinh \psi \cos \theta - {3\imath \over 4} a (\sinh^2 \psi
+\sin^2 \theta) \eea

In conclusion, we have found an apparently consistent order by
order expansion for a two-parameter family of ${1\over 2}$-BPS
supergravity solutions in an asymptotically $AdS_3 \times S^3$
background.

 In this solution, the
five-form RR field-strength $F$ vanishes, while the 3-form RR
field strength $G$ is turned on.   This is a bit puzzling, because
we began with the $\kappa$-symmetry projector for a D3-brane
wrapped along an $AdS_2 \times S^2$ submanifold, and imposed that
projector on the Killing spinors.  This would seem to suggest that
the source is wrapped on an $S^2$ of vanishing size (ie., a sphere
wrapping the $\phi$, $\chi$ coordinates, but fixed at $\theta
=0,\pi$). For the case $W=1$, the solution can be found exactly
\cite{Kumar:2002wc} and this is indeed the case.  But for the more
general case found above, the source does not appear to be
localized at a zero-size sphere (at least to first-order).

We believe that the resolution lies in the fact that the location
of the source ($\psi_0$, $\theta_0$) dropped out of the expression
for the $\kappa$-symmetry projector. In other words, supersymmetry
alone will not give us a unique solution, but rather will
include all
${1\over 2}$-BPS solutions that satisfy this projector, even if
they arise from a source different form the D3-brane used in our
analysis.
 It would be very
interesting to resolve these issues, as well to explicitly
construct this solution to all orders.

\vskip .2in

{\bf Acknowledgments} \vskip .1in

We thank M. Duff and K. Intriligator for useful discussions.
The work of J. K. is supported by DOE-FG03-97ER40546.  J. K. gratefully
acknowledges the University of California, San Diego, where much
of this work was completed.

\end{document}